\documentclass[%
 aip,
 amsmath,amssymb,
 reprint,%
]{revtex4-1}

\usepackage{graphicx}
\usepackage{dcolumn}
\usepackage{bm}

\usepackage[utf8]{inputenc}
\usepackage[T1]{fontenc}
\usepackage{mathptmx}
\usepackage{booktabs}
\usepackage{threeparttable}

\begin{document}


\title{
Contrasted Sn Substitution effects \\
on Dirac line node semimetals $\text{SrIrO}_{3}$ and $\text{CaIrO}_{3}$
} 



\author{Masamichi Negishi}
\affiliation{
Department of Physics, University of Tokyo, 
7-3-1 Hongo, Bunkyo-ku, Tokyo 113-0033, Japan}

\author{Naoka Hiraoka}
\affiliation{
Department of Physics, University of Tokyo, 
7-3-1 Hongo, Bunkyo-ku, Tokyo 113-0033, Japan}

\author{Daisuke Nishio-Hamane}
\affiliation{
Institute for Solid State Physics, University of Tokyo, 
5-1-5 Kashiwanoha, Kashiwa, Chiba 277-8581, Japan}

\author{Hidenori Takagi}
\email[E-mail: ]{h.takagi@fkf.mpg.de}
\affiliation{
Department of Physics, University of Tokyo, 
7-3-1 Hongo, Bunkyo-ku, Tokyo 113-0033, Japan}
\affiliation{
Max Planck Institute for Solid State Research, 
Heisenbergstra{\ss}e 1, 70569 Stuttgart, Germany}
\affiliation{
Institute for Functional Matter and Quantum Technologies, University of Stuttgart, 
Pfaffenwaldring 57, 70550 Stuttgart, Germany}


\date{\today}

\begin{abstract}

Perovskite-type iridates $\text{SrIrO}_{3}$ and $\text{CaIrO}_{3}$ are a Dirac line node semimetal protected by crystalline symmetry, providing an interesting playground to investigate electron correlation effects on topological semimetals. The effect of Sn doping was examined by growing $\text{SrIr}_{1-x}\text{Sn}_{x}\text{O}_{3}$ and $\text{CaIr}_{1-x}\text{Sn}_{x}\text{O}_{3}$ thin films epitaxially on $\text{SrTiO}_{3}$(001) substrate using pulsed laser deposition. Upon Sn doping, the semimetallic ground state switches into an insulator. As temperature is lowered, the resistivity, $\rho (T)$, of $\text{SrIr}_{1-x}\text{Sn}_{x}\text{O}_{3}$ above a critical doping level ($x_{\text{c}} \sim 0.1$) shows a well-defined transition from the semimetal to a weakly ferromagnetic insulator at $T = T_{\text{c}}$. In contrast, the $\rho (T)$ of $\text{CaIr}_{1-x}\text{Sn}_{x}\text{O}_{3}$ with increasing $x$ shows a rapid increase of magnitude but does not show clear signature of metal-insulator transition in the temperature dependence. We argue that the contrasted behavior of the two closely related iridates reflects the interplay between the effects of electron correlation and disorder enhanced by Sn doping.

\end{abstract}


\maketitle 


Recently 5\textit{d} iridium oxides with perovskite-related structures have been explored extensively as a mine for exotic quantum phases, partly because of an interplay of strong spin-orbit interaction and electron correlation of the 5\textit{d} electrons~\cite{Witczak-KrempaARCMP2014}. The strong spin-orbit coupling of $\sim 0.4~\text{eV}$ for 5\textit{d} electrons, which is larger than the typical crystal field splitting of $\lesssim 0.1~\text{eV}$ within $t_{2g}$~\cite{KatukuriPRB2012}, splits the $t_{2g}$ bands with five \textit{d} electrons into upper half-filled $J_{\text{eff}} = 1/2$ band and lower completely filled $J_{\text{eff}} = 3/2$ bands. In the two-dimensional layered perovskite $\text{Sr}_{2}\text{IrO}_{4}$, a modest on-site Coulomb $U$ of $\sim 2~\text{eV}$~\cite{MoonPRL2008} brings the system to a spin-orbital Mott state with the $J_{\text{eff}} = 1/2$ moments~\cite{KimPRL2008, KimScience2009}. Such a spin-orbital Mott state has been identified in many two-dimensional layered iridium oxides, where exotic magnetism of $J_{\text{eff}} = 1/2$, particularly the Kitaev spin liquid state, has been explored~\cite{TakagiNRP2019}. In the three-dimensional perovskites $\text{SrIrO}_{3}$ and $\text{CaIrO}_{3}$, in contrast, the $J_{\text{eff}} = 1/2$ band remains metallic marginally due to the increased band width~\cite{MoonPRL2008} and forms a “correlated” topological semimetal.

\begin{figure}[t]%
    \includegraphics[width=8.5cm]{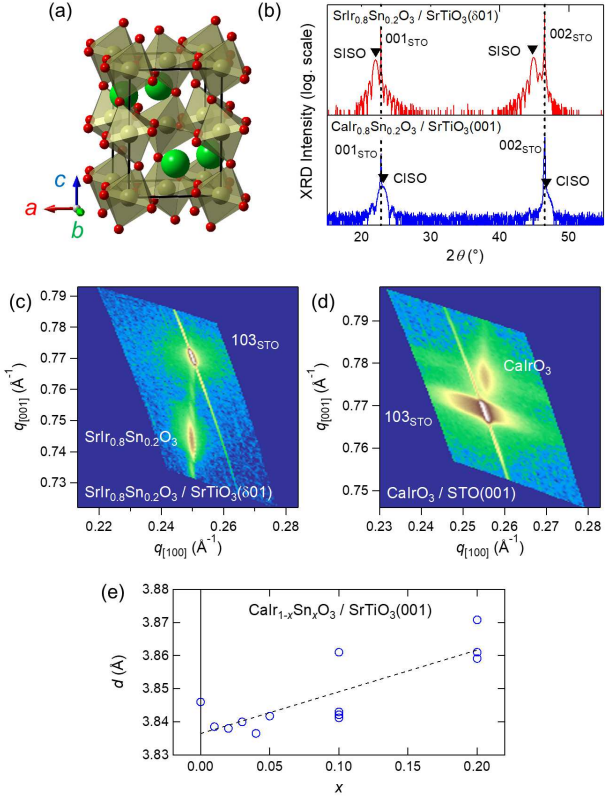}
    \caption{
        (a) Crystal structure of perovskite-type $\text{SrIrO}_{3}$.
        The green, yellow, and red balls indicate Sr, Ir, O atoms, respectively.
        The black box indicates the $\text{GdFeO}_{3}$-type unit cell.
        (b) XRD 2$\theta$-$\theta$ scans of $\text{SrIr}_{0.8}\text{Sn}_{0.2}\text{O}_{3}$ 
        and $\text{CaIr}_{0.8}\text{Sn}_{0.2}\text{O}_{3}$ thin films.
        (c, d) XRD-RSM around $\text{103}_{\text{STO}}$ Bragg peak 
        of $\text{SrIr}_{0.8}\text{Sn}_{0.2}\text{O}_{3}$ and $\text{CaIrO}_{3}$,
        indicating the in-plane lattice constant is locked to the substrate. 
        The lattice constant for $\text{SrIr}_{0.8}\text{Sn}_{0.2}\text{O}_{3}$ 
        along the other in-plane direction $\text{[010]}_{\text{STO}}$ 
        is confirmed to be also fixed to the substrate. 
        (e) The out-of-plane lattice constant $d$ 
        of $\text{CaIr}_{1-x}\text{Sn}_{x}\text{O}_{3}$ films. 
        The black dashed line, the result of linear fitting, 
        shows $\sim 0.7~\%$ increase from $x = 0$ to $0.2$.
        }
    \label{mfig_struct}
\end{figure}%

$\text{SrIrO}_{3}$ and $\text{CaIrO}_{3}$ have a $\text{GdFeO}_{3}$-type distorted perovskite structure where the rotation and the buckling of $\text{IrO}_{6}$ octahedra give rise to a unit cell with the size of $\sqrt{2} a_{\text{c}} \times \sqrt{2} a_{\text{c}} \times 2 a_{\text{c}}$ ($a_{\text{c}}$ is the lattice constant of the original cubic lattice) as shown in Fig.~\ref{mfig_struct}a~\cite{LongoJSSC1971, McDanielJSSC1972}. The results of band calculations indicate that $\text{SrIrO}_{3}$ and $\text{CaIrO}_{3}$ have Dirac electron bands with a line of nodes and heavy hole bands at the Fermi level~\cite{CarterPRB2012}. The line nodes of Dirac bands are protected by the time-reversal symmetry and the gliding symmetry of the $\text{GdFeO}_{3}$-type perovskite structure~\cite{ChenPRB2016}. The existence of Dirac electrons has been supported by experiments, for example, by ARPES~\cite{NiePRL2015, LiuSciRep2016} and transport measurements~\cite{FujiokaPRB2017, FujiokaNatComm2019}. Their proximity to the spin-orbital Mott state and moderately strong electron correlations are evident, for example, from the observation of a transition from a Dirac semimetal to a magnetic insulator by decreasing the number of $\text{SrIrO}_{3}$ layers in $(\text{SrIrO}_{3})_{n}, (\text{SrTiO}_{3})_{1}$ superlattice structures~\cite{MatsunoPRL2015}. The presence of an apparent correlation effect is a prominent feature of the two Ir perovskites when compared with many other topological semimetals which are only weakly correlated, and makes them an interesting arena to explore the effect of electron correlation in topological semimetals. As the ionic radius of $\text{Ca}^{2+}$ ($1.34~\text{\AA}$) is smaller than that of $\text{Sr}^{2+}$ ($1.44~\text{\AA}$)~\cite{ShannonActaCryst1976}, the lattice is more distorted in $\text{CaIrO}_{3}$ than in $\text{SrIrO}_{3}$, which reduces the band width of $\text{CaIrO}_{3}$ appreciably as compared with that of $\text{SrIrO}_{3}$. The effect of electron correlation should be larger in $\text{CaIrO}_{3}$ than $\text{SrIrO}_{3}$. Recently, it was shown that the strong electron correlation modifies the semimetallic band structures appreciably in the two three-dimensional iridium perovskites. The stronger electron correlation effects of $\text{CaIrO}_{3}$ bring its Fermi level closer to the Dirac node and therefore further reduces its density of electrons and holes when compared to $\text{SrIrO}_{3}$~\cite{FujiokaNatComm2019}.

A transition from the Dirac semimetal to a magnetic insulator was also discovered in bulk polycrystalline $\text{SrIrO}_{3}$ by substituting Ir with Sn~\cite{CuiPRL2016}. As Sn ions are tetravalent like Ir ions in $\text{SrIrO}_{3}$, Sn doping does not change the valence of Ir ions and therefore reduces the hopping path of the Ir 5\textit{d} electrons in real space, and hence the effective width of $J_{\text{eff}} = 1/2$ band, as in the case of the superlattice structure. We note here that Sn doping should modify not only the effective band width, but also the degree of disorder. With this unique opportunity of controlling the electron correlations and disorder in mind, we synthesized epitaxial thin films of $\text{SrIr}_{1-x}\text{Sn}_{x}\text{O}_{3}$ and $\text{CaIr}_{1-x}\text{Sn}_{x}\text{O}_{3}$ and measured their resistivity to probe the effects of Sn doping. A transition from a semimetal to a (magnetic) insulator is observed in both $\text{SrIr}_{1-x}\text{Sn}_{x}\text{O}_{3}$ and $\text{CaIr}_{1-x}\text{Sn}_{x}\text{O}_{3}$ thin films, as in the bulk $\text{SrIr}_{1-x}\text{Sn}_{x}\text{O}_{3}$. We discovered a sharp contrast in the transition behavior between the Sr and Ca iridium perovkites: the appearance of well-defined transition from the Dirac semimetal to a magnetic insulator as function of temperature above a critical Sn concentration $x_{\text{c}} \sim 0.1$ for $\text{SrIrO}_{3}$, and the continuous increase of resistivity without a clear signature of the semimetal-insulator transition for $\text{CaIrO}_{3}$, which we ascribe to the interplay of the electron correlation and the disorder effect.

$\text{SrIr}_{1-x}\text{Sn}_{x}\text{O}_{3}$ and $\text{CaIr}_{1-x}\text{Sn}_{x}\text{O}_{3}$ thin films were epitaxially grown in the range of $0 \leq x \leq 0.2$ on $\text{SrTiO}_{3}$(001) substrates by pulsed laser deposition technique using polycrystalline targets with 5~\% excess $B$ (Ir/Sn) cations. The film deposition of $\text{SrIr}_{1-x}\text{Sn}_{x}\text{O}_{3}$ ($\text{CaIr}_{1-x}\text{Sn}_{x}\text{O}_{3}$) was conducted at $650~^{\circ}\text{C}$ ($750~^{\circ}\text{C}$) in a $100~\text{mTorr}$ $\text{O}_{2}$ atmosphere. Typical thickness of our samples estimated from X-ray reflectivity measurement is $\sim 15~\text{nm}$ ($10~\text{nm}$). Magnetization and transport measurements were conducted using a commercial SQUID magnetometer (MPMS, Quantum Design) and in a physical property measurement system (PPMS, Quantum Design). The X-ray diffraction (XRD) measurements were performed using SmartLab, Rigaku.

The results of 2$\theta$-$\theta$ scans of XRD indicate that all the grown $\text{SrIr}_{1-x}\text{Sn}_{x}\text{O}_{3}$ and $\text{CaIr}_{1-x}\text{Sn}_{x}\text{O}_{3}$ films crystalize in the perovskite structure without any trace of impurity phase within the given resolution (Fig.~\ref{mfig_struct}b). The full width at half maximum of the rocking curve is typically as narrow as $0.1^{\circ}$ at the pseudo-cubic (001) peak and the Laue oscillations around the Bragg peaks are clearly observed. Reciprocal space mapping (RSM) measurements reveal that the in-plane lattice of the films is locked to that of the $\text{SrTiO}_{3}$ substrate (Figs.~\ref{mfig_struct}c, d). Those observations clearly demonstrate the epitaxial growth and high crystallinity of the grown films. The out-of-plane lattice constant of the $\text{CaIr}_{1-x}\text{Sn}_{x}\text{O}_{3}$ films as a function of Sn content $x$, as an average, increases almost 0.7~\% from $x = 0$ to $x = 0.2$ which should reflect the larger ionic radius of $\text{Sn}^{4+}$ ($0.690~\text{\AA}$) than that of $\text{Ir}^{4+}$ ($0.625~\text{\AA}$)~\cite{ShannonActaCryst1976} (Fig.~\ref{mfig_struct}e)

\begin{figure*}[t]%
    \includegraphics[width=14cm]{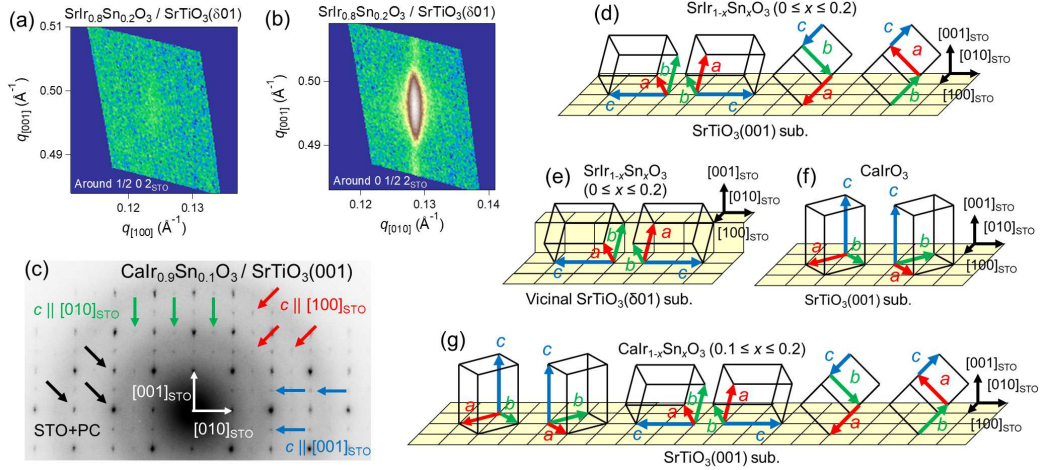}
    \caption{
        The orientation of $\text{SrIr}_{1-x}\text{Sn}_{x}\text{O}_{3}$
        and $\text{CaIr}_{1-x}\text{Sn}_{x}\text{O}_{3}$ films
        by XRD-RSM and transmission electon microscopy.
        XRD-RSM around $\text{1/2~0~2}_{\text{STO}}$ (a) 
        and $\text{0~1/2~2}_{\text{STO}}$ (b)
        for $\text{SrIr}_{0.8}\text{Sn}_{0.2}\text{O}_{3}$ films grown
        on vicinal $\text{SrTiO}_{3}$($\delta$01) substrates. 
        The intensity of film peaks around $\text{0~1/2~2}_{\text{STO}}$
        are significantly stronger than those around $\text{1/2~0~2}_{\text{STO}}$
        (and $\text{0~1~5/2}_{\text{STO}}$, not shown), 
        indicating the preferred orientation of $c$ axis parallel
        to $\text{[010]}_{\text{STO}}$. 
        The comparison of peak intensities of $\text{1/2~0~2}_{\text{STO}}$ 
        and $\text{0~1/2~2}_{\text{STO}}$ peaks normalized with the substrate peaks 
        indicates the ratio of domains with the preferred orientation
        as 95.4~\% for $x = 0$ (data not shown), and 99.0~\% for $x = 0.2$, respectively. 
        (c) The electron diffraction pattern 
        of $\text{CaIr}_{0.9}\text{Sn}_{0.1}\text{O}_{3}$, 
        indicating the coexistence of domains with the doubled $c$ axis 
        perpendicular to the plane as observed in $\text{CaIrO}_{3}$ 
        and those with the doubled $c$ axis lying in the plane (green and red). 
        Schematic pictures of crystalline orientations 
        of $\text{SrIr}_{1-x}\text{Sn}_{x}\text{O}_{3}$ 
        and $\text{CaIr}_{1-x}\text{Sn}_{x}\text{O}_{3}$ films 
        with respect to $\text{SrTiO}_{3}$(001) substrate. 
        $\text{SrIr}_{1-x}\text{Sn}_{x}\text{O}_{3}$ on $\text{SrTiO}_{3}$(001) (d) 
        and on vicinal $\text{SrTiO}_{3}$($\delta$01) (e). 
        $\text{CaIr}_{1-x}\text{Sn}_{x}\text{O}_{3}$ with $x \sim 0$ (f) 
        and $x \geq 0.1$ (g). 
        The black boxes and the arrows labeled by $a$, $b$ and $c$ 
        indicate the bulk unit cells of the film layers.
        }
    \label{mfig_orientation}
\end{figure*}%

\begin{table}[t]%
    \caption{
        The bulk lattice parameters of $\text{SrIrO}_{3}$ and $\text{CaIrO}_{3}$
        and their matching to the cubic lattice parameter of $\text{SrTiO}_{3}$
        ($a_{\text{STO}} = 3.905~\text{\AA}$).
        }
    \label{table_latconst}
    \begin{tabular}{cccccc} \toprule
        & $a$ & $b$ & $c$
        & $\frac{a_{\text{pc}}}{a_{\text{STO}}} - 1$
        & $\frac{c_{\text{pc}}}{a_{\text{STO}}} - 1$ \\ \midrule
        $\text{SrIrO}_{3}$~\cite{ChamberlandJAC1992}
        & $5.597(1)~\text{\AA}$ & $5.581(1)~\text{\AA}$ & $7.752(2)~\text{\AA}$
        & $+ 1.2~\%$ & $- 0.7~\%$ \\
        $\text{CaIrO}_{3}$~\cite{FujiokaNatComm2019}
        & $5.3597(5)~\text{\AA}$ & $5.6131(4)~\text{\AA}$ & $7.6824(8)~\text{\AA}$
        & $- 0.6~\%$ & $- 1.6~\%$ \\ \bottomrule
    \end{tabular}
\end{table}%

The film orientation was identified by RSM measurements of XRD and transition electron microscopy (Figs.~\ref{mfig_orientation}a-c). We will describe the lattice orientation of films by the pseudo-cubic unit cell in this paper using lattice parameters $a_{\text{pc}} = \sqrt{a^{2} + b^{2}} / 2$ and $c_{\text{pc}} = c/2$, where $a$, $b$ and $c$ denote the orthorhombic unit cell parameters of the distorted $\text{GdFeO}_{3}$ structure~\cite{OrthoMono}. In case of $\text{SrIr}_{1-x}\text{Sn}_{x}\text{O}_{3}$, the $c_{\text{pc}}$ axis (|| orthorhombic $c$) lies within the substrate plane independent of Sn doping (Fig.~\ref{mfig_orientation}d). This is natural because $c_{\text{pc}}$ is closer to that of $\text{SrTiO}_{3}$ substrate than $a_{\text{pc}}$ (See Table~\ref{table_latconst}.). In contrast, $a_{\text{pc}}$ is closer to that of $\text{SrTiO}_{3}$ in $\text{CaIrO}_{3}$. The $c_{\text{pc}}$ axis (|| orthorhombic $c$) therefore aligns perpendicular to the substrate plane in the $\text{CaIrO}_{3}$ case (Fig.~\ref{mfig_orientation}f)~\cite{HiraiAPL2015}. The inclusion of minority domains with the $c_{\text{pc}}$ axis lying within the substrate plane as in $\text{SrIr}_{1-x}\text{Sn}_{x}\text{O}_{3}$ are observed for the high Sn content films ($x \geq 0.1$) (Fig.~\ref{mfig_orientation}g), which very likely reflects the expansion of $c_{\text{pc}}$ due to Sn doping and the resultant proximity to the lattice constant of $\text{SrTiO}_{3}$.

\begin{figure}[t]%
    \includegraphics[width=8.5cm]{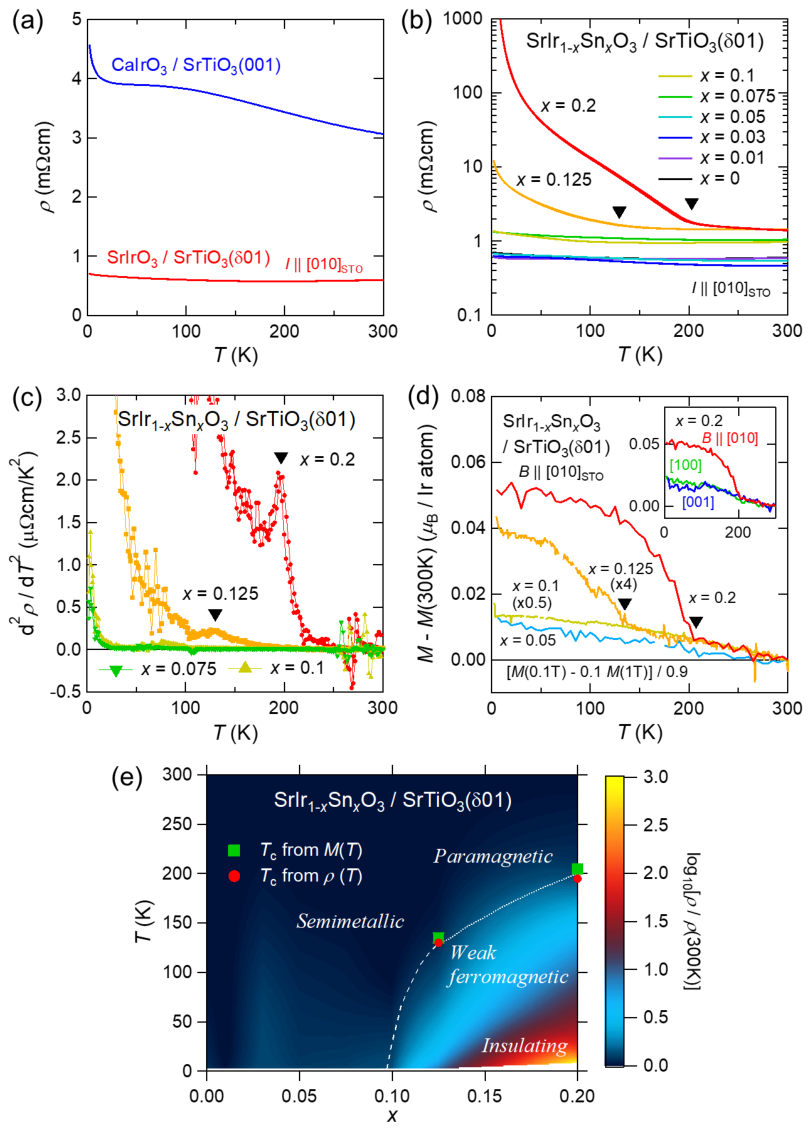}
    \caption{
        (a) Temperature-dependent resistivity $\rho (T)$ 
        of $\text{SrIrO}_{3}$ and $\text{CaIrO}_{3}$. 
        (b) $\rho (T)$ of $\text{SrIr}_{1-x}\text{Sn}_{x}\text{O}_{3}$ 
        and (c) their second derivative. 
        (d) Temperature-dependent magnetization $M (T)$ 
        of $\text{SrIr}_{1-x}\text{Sn}_{x}\text{O}_{3}$ 
        measured in the external field $B~||~\text{[010]}_{\text{STO}}$. 
        The background contributions including that from the contaminated oxygen 
        was subtracted by taking the difference 
        between $0.1~\text{T}$ and $1~\text{T}$ data 
        as $M(T) = [M(T, B = 0.1~\text{T}) - 0.1 M(T, B = 1~\text{T})] / 0.9$. 
        A weak ferromagnetic moment was not observed 
        for the other magnetic field directions
        $B \ || \ \text{[100]}$ and [001]
        as shown in the inset.
        (e) The weak ferromagnetic transition temperature (the green squares) 
        and the temperature where the resistivity has the anomaly (the red circles). 
        The color plot shows the ratio of $\rho (T)$ and $\rho (T = 300~\text{K})$ 
        in logarithmic scale.
        }
    \label{mfig_rtsiso}
\end{figure}

As $a_{\text{pc}}$ is larger than $c_{\text{pc}}$, the orientation of $c_{\text{pc}}$ within the substrate plane can be controlled by introducing additional epitaxial strain using step edges of vicinal substrate~\cite{ZhangASS2013, JaiswalArXiv2019}. We used a vicinal $\text{SrTiO}_{3}$($\delta$01) substrate with the substrate plane $0.4^{\circ}$ rotated from (001) towards the [100] direction for the growth of $\text{SrIr}_{1-x}\text{Sn}_{x}\text{O}_{3}$. Because of the epitaxial strain from the side (100) plane at the step edges, the $c_{\text{pc}}$ (|| orthorhombic $c$) axis prefers to align along the edge, namely $\text{[010]}_{\text{STO}}$ direction (Fig.~\ref{mfig_orientation}e). The RSM measurements clearly indicate that more than 95~\% of domains have $c_{\text{pc}}$ axis parallel to the substrate [010] direction for the films on the vicinal substrates. Consistent with the in-plane preferred orientation, a clear anisotropy of magnetization within the substrate plane was observed for $x = 0.2$ sample on the vicinal substrate as we describe below. Pronounced anisotropy in the magnetization was not observed in the resistivity $\rho (T)$.

The resistivity $\rho (T)$ measurements on the $\text{SrIr}_{1-x}\text{Sn}_{x}\text{O}_{3}$ films indicate the presence of a metal-insulator transition accompanied by a weak ferromagnetism as in the bulk~\cite{CuiPRL2016}. The $\text{SrIrO}_{3}$ ($x = 0$) film shows only weakly temperature-dependent behavior of $\rho (T)$, where a gradual increase followed by the temperature-independent behavior is observed with decreasing temperature (Fig.~\ref{mfig_rtsiso}a). This agrees well with previous reports on $\text{SrIrO}_{3}$ thin films~\cite{MatsunoPRL2015, ZhangCRSSMS2018} and can be understood as a typical behavior of semimetals with extra conductivity at high temperature from thermally excited electrons and holes. With Sn doping (Fig.~\ref{mfig_rtsiso}b), we do not observe an appreciable change of the $\rho (T)$ up to the critical concentration $x_{\text{c}} = 0.1$. Above $x_{\text{c}} = 0.1$, however, the resistivity shows a transition from the semimetal to a weak insulator at a transition temperature $T_{\text{c}}$ where we observe a kink in $\rho (T)$ and a well-defined peak of the second derivative $\mathrm{d}^{2} \rho (T) / \mathrm{d} T^{2}$ (Fig.~\ref{mfig_rtsiso}c). $T_{\text{c}}$ increases rapidly with increasing $x$. The metal-insulator transition below $T_{\text{c}}$ is accompanied by a weak ferromagnetism with the easy axis parallel to the pseudo-cubic $c_{\text{pc}}$ axis (|| orthorhombic $c$), as seen in Fig.~\ref{mfig_rtsiso}d. This behavior, the emergence of the magnetic insulator out of the Dirac node semimetal with Sn doping, can be summarized as a Sn content $x$-$T$ phase diagram on top of contour map of the magnitude of $\rho (T)$ (Fig.~\ref{mfig_rtsiso}e). The stabilization of the magnetic insulator phase at low temperatures above $x_{\text{c}} = 0.1$ highly likely originates from the increase of the effective electron correlation. We argue that this is because of the reduced hopping of $J_{\text{eff}} = 1/2$ electrons by the introduction of $\text{Sn}^{4+}$ without conduction electrons.

We find contrasting behavior in the Sn doping effect on $\rho (T)$ in $\text{CaIr}_{1-x}\text{Sn}_{x}\text{O}_{3}$ compared to $\text{SrIr}_{1-x}\text{Sn}_{x}\text{O}_{3}$. For $\text{CaIrO}_{3}$ ($x = 0$) films (Fig.~\ref{mfig_rtsiso}a), the overall behavior of $\rho (T)$ is similar to that observed in $\text{SrIrO}_{3}$. The magnitude of resistivity is, however, appreciably larger than that of $\text{SrIrO}_{3}$. A weak increase of $\rho (T)$, reminiscent of a weak localization and not observed in $\text{SrIrO}_{3}$, is seen below $20~\text{K}$, which is suggestive of the presence of appreciable disorder effect. It was discussed in the recent transport study on single crystal $\text{CaIrO}_{3}$ that the Fermi level is much closer to the Dirac nodes and hence the electron and the hole densities are lower in $\text{CaIrO}_{3}$ than $\text{SrIrO}_{3}$ due to the enhanced correlation effect originating from the narrow band in $\text{CaIrO}_{3}$~\cite{FujiokaNatComm2019}. The effect of disorder should be enhanced in $\text{CaIrO}_{3}$ because of the reduced carrier density, which may account for the larger resistivity and the weakly localized behavior.

\begin{figure}[t]%
    \includegraphics[width=8.5cm]{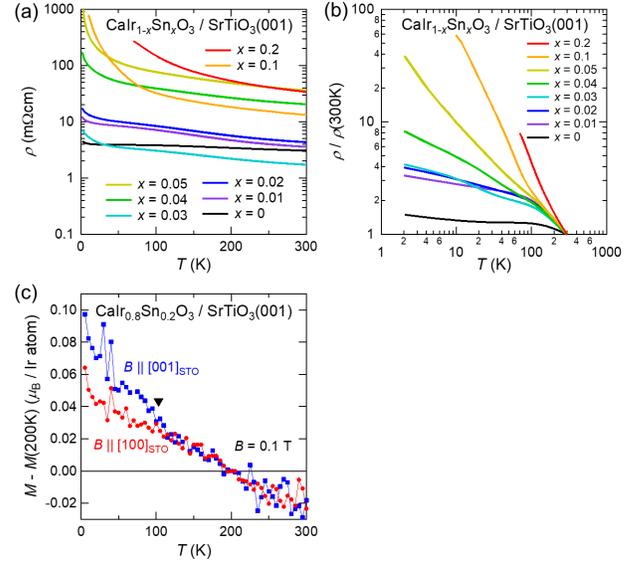}
    \caption{
        (a) Temperature-dependent resistivity $\rho (T)$ 
        of $\text{CaIr}_{1-x}\text{Sn}_{x}\text{O}_{3}$ thin films 
        and (b) the ratio of $\rho (T)$ in (a) and $\rho (T = 300~\text{K})$,
        indicating power-law temperature dependence.
        (c) Temperature-dependent magnetization $M(T)$
        of $\text{CaIr}_{0.8}\text{Sn}_{0.2}\text{O}_{3}$ thin film,
        plotted as the difference from $M ( T = 200~\text{K})$.
        An additional contribution to the uncompensated offset
        can be seen below $T \sim 100~\text{K}$ for $B \ || \ \text{[001]}_{\text{STO}}$.
        }
    \label{mfig_rtciso}
\end{figure}%

With Sn substitution, $\rho (T)$ gradually increases and shows a poorly insulating behavior with a power-law divergence (Figs.~\ref{mfig_rtciso}a, b). We do not see a well-defined semimetal-insulator transition as a function of $T$ and $x$ in contrast to the case for $\text{SrIrO}_{3}$. We argue that the gradual transition from the Dirac node semimetal to a weak insulator is driven by the disorder and perhaps the inhomogeneity, and that the nature of semimetal-insulator transition is distinct from that of $\text{SrIrO}_{3}$. It is natural that the effect of disorder associated with Sn-doping is much more profound in $\text{CaIrO}_{3}$ than in $\text{SrIrO}_{3}$ because of the lower carrier density and Fermi energy of $\text{CaIrO}_{3}$. For $\text{CaIr}_{0.8}\text{Sn}_{0.2}\text{O}_{3}$, a very weak ferromagnetic moment appears to emerge below $T_{\text{mag}} \sim 100 \ \text{K}$ with $B$ perpendicular to the film plane, smaller in magnitude and lower in temperature than $\text{SrIr}_{0.8}\text{Sn}_{0.2}\text{O}_{3}$ as shown in Fig.~\ref{mfig_rtciso}c. No clear anomaly can be identified in $\rho (T)$ at $T_{\text{mag}}$ and $\rho (T)$ shows an insulating behavior above $T_{\text{mag}}$, implying that the magnetic ordering is not a trigger of the semimetal-insulator transition in $\text{CaIr}_{1-x}\text{Sn}_{x}\text{O}_{3}$ in contrast to that in $\text{SrIr}_{1-x}\text{Sn}_{x}\text{O}_{3}$. The emergence of magnetism in the disordered insulator may suggest a Mott insulator character and therefore a Mott-Anderson type metal-insulator transition in $\text{CaIr}_{1-x}\text{Sn}_{x}\text{O}_{3}$. 

In summary, we have successfully grown thin films of Dirac semimetals $\text{SrIr}_{1-x}\text{Sn}_{x}\text{O}_{3}$ and $\text{CaIr}_{1-x}\text{Sn}_{x}\text{O}_{3}$ epitaxially on $\text{SrTiO}_{3}$(001), with their orthorhombic $c$ axis parallel and perpendicular to the substrate plane respectively. In the case of $\text{SrIr}_{1-x}\text{Sn}_{x}\text{O}_{3}$, the $c$ axis can be aligned within the substrate plane using a vicinal substrate. While a well-defined Dirac node semimetal to a magnetic insulator transition with Sn doping is observed at a critical Sn concentration $x_{\text{c}} \sim 0.1$ in $\text{SrIr}_{1-x}\text{Sn}_{x}\text{O}_{3}$, we observe in $\text{CaIr}_{1-x}\text{Sn}_{x}\text{O}_{3}$ that the Dirac node semimetal changes only gradually to a poor insulator without any well-defined semimetal-insulator transition. We argue that the contrast between $\text{SrIr}_{1-x}\text{Sn}_{x}\text{O}_{3}$ and $\text{CaIr}_{1-x}\text{Sn}_{x}\text{O}_{3}$ is a consequence of the interplay between the enhanced effective electron correlation and the disorder effect by Sn doping. While the correlation effect dominates in the case of $\text{SrIr}_{1-x}\text{Sn}_{x}\text{O}_{3}$, the disorder effect and the carrier localization dominate in the $\text{CaIr}_{1-x}\text{Sn}_{x}\text{O}_{3}$ due to the proximity of the Fermi level to the Dirac nodes and the resultant low carrier concentration. These results clearly indicate that Sn-doped iridium perovskite oxides are an interesting playground to study the effect of electron correlations and disorders in a topological semimetal.


%
%

%


\begin{acknowledgments}
This work was supported
by JSPS KAKENHI Grant Number JP24224010, JP17H01140, JP15H06092, JP17K14335 and JP18J21922.
\end{acknowledgments}

\bibliography{ref.bib}

\end{document}